\documentclass[12pt]{iopart}

\def\Grad{\mbox{\boldmath$\nabla$}}

\def\um{\mu\mbox{m}}
\def\sec{\mbox{s}}
\def\psec{\mbox{ps}}

\def\Wcm2{\mbox{W cm}^{-2}}
\def\Wcmum2{\mbox{Wcm}^{-2}\mu\mbox{m}^{2}}
\def\cm3{\mbox{cm}^{-3}}
\def\Vcm{\mbox{V cm}^{-1}}
\def\lD{\lambda_{\mbox{\tiny D}}}

\def\beq{\begin{equation}}
\def\eeq{\end{equation}}
\def\bea{\begin{eqnarray}}
\def\eea{\end{eqnarray}}
\def\bi{\begin{itemize}}
\def\ei{\end{itemize}}
\def\bp{\begin{picture}}
\def\ep{\end{picture}}
\def\bc{\begin{center}}
\def\ec{\end{center}}

\usepackage{iopams}  

\usepackage{graphicx}
\usepackage{color}

\usepackage{amssymb}
\usepackage{hyperref}
\usepackage{dcolumn}
\usepackage{bm}

\begin{document}

\title[Electric field dynamics and ion acceleration]{Electric field dynamics and ion acceleration in the self-channeling of a superintense laser pulse}

\author{Andrea Macchi$^{1,2}$, Francesco Ceccherini$^2$, Fulvio Cornolti$^2$,
Satyabrata Kar$^3$, and Marco Borghesi$^3$
}
\address{$^1$polyLAB, CNR-INFM, Pisa, Italy}
\address{$^2$Dipartimento di Fisica ``Enrico Fermi'', Universit\`a di Pisa, Largo Bruno Pontecorvo 3, I-56127 Pisa, Italy}
\address{$^3$School of Mathematics and Physics, the Queen's University of Belfast, Belfast BT7 1NN, UK}

\begin{abstract}
The dynamics of electric field generation and 
radial acceleration of ions by a laser pulse of relativistic intensity
propagating in an underdense plasma 
has been investigated using an one-dimensional 
electrostatic, ponderomotive model developed to 
interpret experimental measurements of electric fields 
[S. Kar et al, New J. Phys. \textbf{9}, 402 (2007)].
Ions are spatially focused at the edge of the 
charge-displacement channel, leading to hydrodynamical breaking, 
which in turns causes the heating of electrons and an ``echo'' effect in 
the electric field. The onset of complete electron depletion in the
central region of the channel leads to a smooth  
transition to a ``Coulomb explosion'' regime and a saturation
of ion acceleration.
\end{abstract}


\submitto{\PPCF}

\maketitle

\section{Introduction}

Self-focusing and guiding of laser pulses is one of the most 
peculiar effects in nonlinear optics. In the case of superintense
laser pulses (having irradiances beyond $10^{18}~\Wcmum2$) propagating
in a plasma, self-focusing arises due to the combined effects of 
the intensity dependence of the index of refraction for relativistic 
electron velocities and of the expulsion of plasma from the propagation axis 
driven by the radial ponderomotive force 
\cite{sunPF87,moriPRL88,esareyIEEE97};
this latter effect leads to self-channeling, i.e. to the creation of 
a low-density channel that may act as a waveguide for the laser pulse.  
As the ponderomotive force acts on the plasma electrons, the channel is
first drilled in the electron density only and becomes strongly charged
\cite{borisovPRL92,borisovPRA92};
the electric field in the channel leads to ion acceleration 
\cite{sarkisovJETP97,sarkisovPRE99,krushelnickPRL99}, which is of interest
both as a diagnostic of the interaction and as a way to provide fast ions
for applications, e.g. for the production of fusion neutrons 
\cite{pretzlerPRE98,golovizninJPD98,fritzlerPRL02}. Besides these applications, the self-channeling process
may be of relevance for laser-plasma acceleration of electron 
\cite{malkaPPCF05} 
and ion beams \cite{weiPRL04,willingalePRL06},
X- and $\gamma$-ray sources \cite{burnettIEEE90,roussePRL04}, 
and fast ignition in
laser-driven Inertial Confinement Fusion
\cite{tabakPP94}.

To interpret experimental results on self-channeling and related 
ion acceleration or neutron production, 
a simple modeling of the radial dynamics based on 
the ponderomotive force and electrostatic field only has often been used
\cite{krushelnickPRL97,krushelnickPRL99,sarkisovJETP97,sarkisovPRE99,golovizninJPD98,karNJP07}.
Such a ponderomotive, electrostatic model (PEM) is attractive due to its
simplicity and easy numerical implementation with respect to multi-dimensional,
electromagnetic particle-in-cell (PIC) simulations which typically require
massively parallel computing, while simulations based on the PEM in one 
spatial dimension (1D), taking only the radial dynamics into account,
can be performed on a personal computer. Of course the simple PEM gives a 
strongly simplified description of the interaction where effects such as the 
nonlinear evolution of the pulse due to self-focusing or instabilities are not
included. The use of the PEM can be justified only by its ability to reproduce,
at least qualitatively, experimental observations for particular regimes.
 
Recently, the proton imaging technique \cite{borghesiRSI03}
allowed for the first time the detection, with spatial and temporal 
resolution, of the electrostatic, slowly-varying fields produced during and 
after the self-channeling process (previous experimental investigations 
\cite{borisovPRL92,monotPRL95,borghesiPRL97,krushelnickPRL97,fedosejevsPRE97,sarkisovJETP97}
were based on optical diagnostics or indirect measurements not directly 
sensitive to electric fields).  
The proton imaging data were well reproduced numerically
by simulating the proton probe deflection in the  
space- and time-dependent  electric field distribution
obtained from PIC simulations based on the 1D PEM. Moreover, a similar dynamics
of the electric field and the ion motion in radial direction was also observed
in two-dimensional (2D) electromagnetic PIC 
simulations of the laser-plasma interaction \cite{karNJP07,macchiPPCF07}.

Motivated by the fair agreement with experimental results and more complex
simulations, we have used the PEM to gain an insight of the radial  
dynamics during and after the self-channeling process. Although there may
be much additional physics at play that is not included in the simple PEM, 
a study based on the latter can be useful to discriminate purely 
ponderomotive and electrostatic effects from those due to other 
contributions,
e.g. self-generated magnetic fields \cite{borghesiPRL97}, anisotropy
and polarization effects \cite{naumovaPRE02}, hosing instabilities
\cite{shvetsPRL94,naumovaPP01}, and so on. 

It turns out that the dynamics contained in the 1D PEM is already quite rich.
We focus our attention on the regime where there is not a complete depletion 
of electrons in the channel and the electric field almost exactly balances the 
ponderomotive force (PF) locally. From the point of view of ion acceleration
we call this the ``ponderomotive'' regime, since the force on ions
is proportional to the PF, while we call ``Coulomb explosion'' the regime 
in which, due to complete electron depletion, the ions move under the
action of their own space-charge field only. Our analysis shows that the
transition between the two regimes occurs rather smoothly.

A prominent effect observed in the simulation is the
spatial ``focusing'' of the ions at the edge of the channel, where 
they form a very sharp spike of the ion density. 
The density spike splits up rapidly due to 
hydrodynamical breaking, and a short bunch of ``fast'' ions is generated.
Density spiking and breaking occur even for pulse durations shorter
than the time needed for ions to reach the breaking point.
The onset of hydrodynamical breaking also causes a strong heating of
electrons and the formation of an ambipolar sheath field around the
breaking point. For pulse durations shorter than the time at which breaking
occurs, there is a sort of ``echo'' effect
in the electric field, which vanishes at the end of the laser pulse
to re-appear later at the breaking location.
Simple analytical descriptions of the spatial focusing mechanism and
of the ambipolar field structure around the density spike are given.
Finally, the smooth transition toward the ``Coulomb explosion'' regime
is described both analytically and via simulations.

\section{The model}

We now give a detailed description of the 1D, electrostatic, ponderomotive
model which has been already used in Ref.\cite{karNJP07} to simulate
the radial electron and ion dynamics due to self-channeling of an intense 
laser pulse in an underdense plasma. 
Only the slowly-varying dynamics of the plasma electrons is taken into 
account, i.e. a temporal average over oscillations at the laser frequency
is assumed. In other words, what we describe is the dynamics of 
electron ``guiding centers'', i.e. of quasi-particles
moving under the action of the ponderomotive force (PF). 
The PF associated to a laser pulse described by the vector potential
${\bf A}({\bf x},t)$ can be written, under suitable conditions
(see e.g. \cite{bauerPRL95}), as 
\bea
{\bf F}_p({\bf x},t)&=&-m_e c^2\Grad\gamma({\bf x},t),\\
\gamma({\bf x},t)&=&\left[1+\langle{\bf a}^2({\bf x},t)\rangle\right],\\
{\bf a}({\bf x},t)&=&(e/m_e c^2){\bf A}({\bf x},t),
\eea 
where the angular brackets denote average over a period.
We assume a non-evolving laser pulse (neglecting pulse diffraction, 
self-focusing and energy depletion) and
complete cylindrical symmetry around the propagation axis,
taking only the radial dynamics into account.
Under such simplifying assumptions, the laser pulse is completely defined 
by the cycle-averaged squared modulus of the vector potential in 
dimensionless units, which we write as
\begin{equation}
a^2(r,t)=\langle{\bf a}^2({\bf x},t)\rangle
      =a^2(r){\cal P}(t)=a^2_0 e^{-r^2/r_0^2}{\cal P}(t).
\label{eq:a2_rt}
\end{equation}
where ${\cal P}(t)$ is the temporal envelope.
The radial component of the PF can thus be written as
\begin{eqnarray}
F_r &=& F_r(r,t)=-m_e c^2 \partial_r \gamma , \\
\gamma &=& \gamma(r,t)
       =\left[{1+{a^2(r,t)}}\right]^{1/2}.
\end{eqnarray}

Unless $a_0 \ll 1$ the electron dynamics is relativistic.  
For $a_0 \gtrsim 1$, besides using the relativistic expression of the
PF one has to account for the inertia due to the high-frequency quiver
motion. This is included via an effective,
position-dependent mass $m=m_e\gamma$ of the quasi-particles \cite{bauerPRL95}.
We thus write for the radial momentum
\begin{equation}
p_{e,r} \simeq m_e\gamma v_r .
\end{equation}
Therefore, the equation of motion for the electrons
is written as 
\begin{equation}
\frac{dp_{e,r}}{dt}=F_r-eE_r
\label{eq:elecmotion}
\end{equation}
where $E_r$ is the electrostatic field due to the space-charge 
displacement. The effect of the laser force on ions
having mass $m_i=Am_p \gg m_e$ can be neglected, leaving 
the electrostatic force $Ze E_r$ as the only force on the ions. 
The ion equation of motion is thus written as
\begin{equation}
\frac{dv_i}{dt}=\frac{Z}{A}\frac{e}{m_p}E_r .
\label{eq:ionmotion}
\end{equation}
The electrostatic field is computed via Poisson's equation
\begin{equation}
\partial_r E_r =4\pi e(Zn_i-n_e) .
\label{eq:poisson}
\end{equation}
Equations (\ref{eq:elecmotion}), (\ref{eq:ionmotion}) and 
(\ref{eq:poisson}) are the basis of our particle simulations
in 1D cylindrical geometry.

Thanks to the low dimensionality of our model, it is possible to use a
very high resolution in the simulations. A typical run used 
$40000$ spatial gridpoints, with a spatial resolution 
$\Delta r=d_p/500$ where $d_p=c/\omega_p$ is the plasma frequency,
and up to $3\times 10^7$ particles for both electron and ion
distributions. It turned out that such a high resolution is needed to resolve
the very sharp spatial structures that are generated during the
simulation, as it will be discussed below. 
The initial temperature of the plasma
is taken to be zero and no significant numerical self-heating is observed 
during the simulations.

In the simulation results shown below, 
the spatial coordinate $r$ is normalized to the laser wavelength $\lambda$, 
the time to the laser period $T_L$,
the density to the critical density $n_c$, 
and the electric field to the ``relativistic'' threshold field $E_0$.
The definition of these parameters and their value in ``practical'' units 
for the typical choice $\lambda=1~\um$ are as follows:
\begin{eqnarray}
T_L&=&\frac{\lambda}{c}=3.34 \times 10^{-15}~\sec,\\
n_c&=&\frac{m_e c^2}{\pi e^2\lambda^2}=1.11 \times 10^{21}~\cm3,\\
E_0&=&2\pi\frac{m_e c^2}{e\lambda}=3.21 \times 10^{10}~\Vcm .
\end{eqnarray}
Momenta are given in units of $m_i c$ for ions and $m_e c$ for electrons,
respectively.

To compare with experiments, we note that the pulse intensity
as a function of time and position is given by
\beq
I(r,t)=\frac{\pi c}{\lambda^2}\left\langle A^2(r,t)\right\rangle
      =\pi\left(\frac{m_e c^2}{e\lambda}\right)^2 a^2(r,t).
\eeq
The relation 
\beq
a_0=0.85 \times \left(\frac{I_0\lambda^2}{10^{18}~\mbox{W cm}^{-2}\mu\mbox{m}^2}\right)^{1/2}
\label{eq:a0}
\eeq
thus gives the parameter $a_0$ as a function of the \emph{maximum} intensity
$I_0$, i.e. as the value of the intensity at the center of the laser spot
and at the pulse peak. Some care should be taken when comparing to experiments 
where some \emph{average} value, i.e. the pulse energy over the
spot area and the pulse duration, is given instead; inserting such an 
average value into Eq.(\ref{eq:a0}) would represent too low a value for a 
proper comparison.

The ``pulse duration'' and ``spot radius'' quoted in experimental papers
most of the times refer to the FWHM of the temporal envelope 
and to half the FWHM of the radial profile of the intensity, respectively.
In our model the
laser pulse intensity varies as
\begin{equation}
{\cal P}(t)=\sin^4\left(\frac{\pi t}{\tau}\right)
           =\sin^4\left(\frac{1.14 t}{\tau_{1/2}}\right).
\end{equation}
for $0<t<\tau$, while ${\cal P}(t)=0$ for $t>\tau$.
(A Gaussian pulse envelope has also been tested, but
no significant differences in the simulation results was evidenced.)
The parameter $\tau_{1/2}=1.14\tau/\pi$ is thus the FWHM 
duration of the laser pulse intensity. For the Gaussian radial profile,
the FWHM radius of the intensity profile
is $r_{1/2}=\sqrt{\ln 2}r_0 \simeq 0.83 r_0$.

\section{Results}

The particle code based on the PEM 
was developed to analyze the experiment of Ref.\cite{karNJP07}, and the most 
relevant results emerged during such analysis. 
Therefore the first simulation we show, which gives us the basis for our 
discussion, has been performed for laser and
plasma parameters in the range of those covered by the experiment of 
Ref.\cite{karNJP07}. In the latter, the laser pulse wavelength
was $\lambda = 1.053~\um$, the intensity was in the range from 
$4 \times 10^{18}$ to $2 \times 10^{19}~\Wcm2$, 
the duration was $\tau_{1/2} \simeq 1.2~\psec \simeq 330~T_L$ and 
the waist radius in vacuum was $r_{1/2} \simeq 5~\lambda$. 
As the effective values of the 
pulse amplitude and radius may change during the propagation into the plasma, 
the values of $a_0$ and $r_0$ were varied in the simulation until the
best match in the reconstruction of proton images was found \cite{karNJP07}.
The plasma was created 
in a gas jet of Helium (charge state $Z=2$, mass number $A=4$) and
the typical electron density $n_e$ was in the range from
$10^{18}$ to $10^{19}~\cm3$, i.e from about $10^{-2}n_c$ to $10^{-1}n_c$.

\subsection{Ion dynamics and hydrodynamical breaking}
\label{sec:breaking}

\begin{figure}
\begin{center}
\includegraphics[angle=0,width=0.8\textwidth]{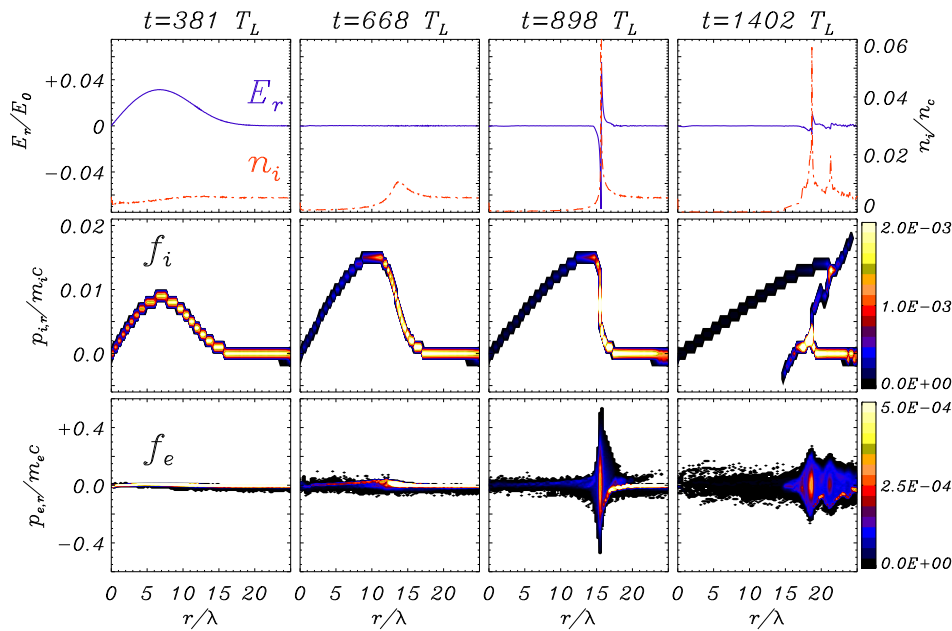}
\end{center}
\caption{(color online) Simulation results in the ponderomotive regime.
The profiles of electric field $E_r$ (blue, thick line)
and ion density $n_i$ (red, dash-dotted line), 
and the phase space distributions
of ions $f_i(r,p_r)$ and electrons $f_e(r,p_r)$
are shown at four instants 
(with times given in units of the laser period $T_L$) 
which are representative of subsequent stages of the dynamics:
the initial ponderomotive acceleration ($t=381T_L$), 
the vanishing of $E_r$ after the end of the laser pulse ($t=668T_L$), 
the spiking of $n_i$ preceding the hydrodynamical breaking 
and the rebirth (``echo'') of $E_r$ correlated with 
strong electron heating ($t=898T_L$), and 
the ``{\sf x}'' structure of the ion phase space after breaking ($t=1402T_L$).
Parameters are 
$a_0=2.7$, $n_e/n_c=10^{-2}$, $r_0=7.5\lambda$, $\tau_{1/2}=330T$.
\label{PIC1D}\label{fig:PAsim}}
\end{figure}

\begin{figure}
\begin{center}
\includegraphics[angle=0,width=0.6\textwidth]{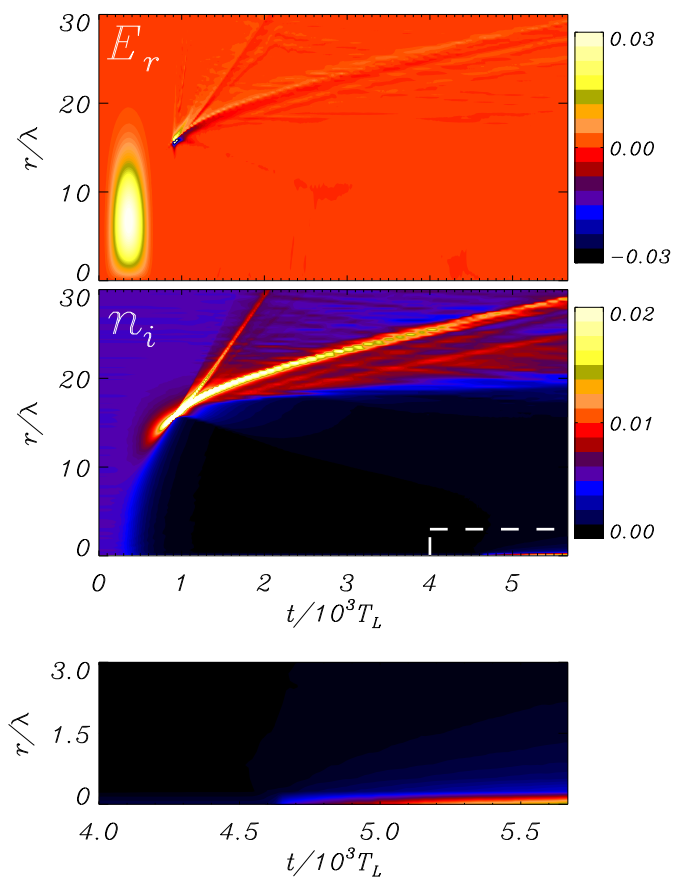}
\end{center}
\caption{(color online) 
Space-time contour plots of $E_r(r,t)$ (top frame)
and $n_i(r,t)$ (middle frame) from the same simulation of
Fig.\ref{fig:PAsim}. 
The bottom frame shows enlarged view of the region 
marked by the dashed contour in the middle frame, 
showing the accumulation of ions on the axis at late times.
The contour levels have been partly saturated to make regions of low
field or density appear.
\label{fig:PAsim_ErtNit}}
\end{figure}

Fig.\ref{fig:PAsim} shows the spatial profiles of the electric field $E_r$ and
of the ion density $n_i$, and the distribution functions in the ($r,p_r$)
phase space of ions ($f_i$) and electrons ($f_e$) at four different times, 
which are representative of the subsequent stages of the dynamics.  
Fig.\ref{fig:PAsim_ErtNit} shows the complete space--time evolution 
of $E_r$ and $n_i$ from the same simulation as contour plots. 
The parameters are $a_0=2.7$, $n_e/n_c=0.01$, $r_0=7.5\lambda$,
$\tau_{1/2}=330T_L$.

The dynamics observed in the simulation of Fig.\ref{fig:PAsim} can be 
described as follows.
During the laser pulse ($t<\tau$),
the PF $F_r$ pushes the electrons outwards,
quickly creating a positively charged channel along the axis. 
This charge displacement creates a radial ES field $E_r$ 
which holds the electrons back, as shown in the first frame of
Fig.\ref{fig:PAsim} for $t=381T_L$ (i.e. $\simeq 50$ periods after the pulse
peak). 
In this stage, we find the ES field to balance almost exactly the
PF, i.e. $eE_r \simeq F_r$ (when plotting $F_r/e$ as well 
in Fig.\ref{fig:PAsim} for $t=381$, its profile cannot be distinguished
from that of $E_r$). 
Thus, at any time the electrons are approximately
in a state of mechanical equilibrium. 
From the electron phase space at $t=381T_L$
we also observe that no significant electron heating 
occurs, $f_e$ being a narrow function along the $p_{e,r}$ axis.
During this stage the ions are accelerated by the electric force $ZeE_r=ZF_r$,
and a depression in $n_i$ is thus produced around the axis. 
After the end of the pulse ($t \simeq 668T_L$ in Fig.\ref{fig:PAsim}), 
ion acceleration is over, the peak momentum of the ions is $0.015m_i c$,
and $E_r\simeq 0$. This indicates that
the electrons have rearranged their spatial distribution in order to
restore the local charge neutrality. At the same time, we still observe 
a very weak heating of electrons, consistently with the keeping of the 
mechanical quasi-equilibrium condition up to this stage.

\begin{figure}
\begin{center}
\includegraphics[angle=0,width=0.6\textwidth]{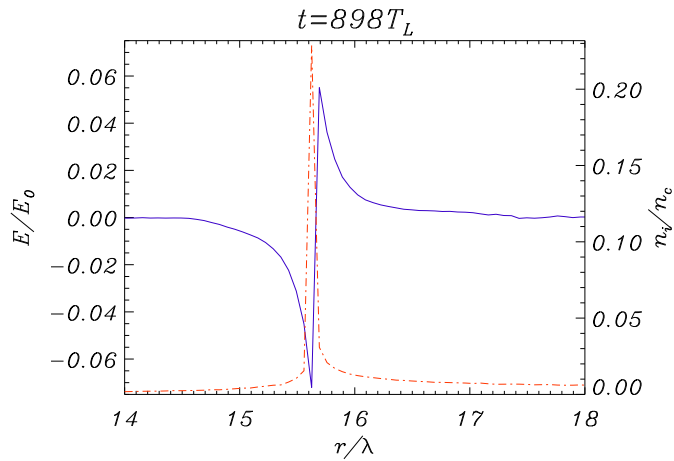}
\end{center}
\caption{(color online) 
Detail of the profiles of electric field $E_r$ (blue, thick line)
and ion density $n_i$ (red, dash-dotted line)
around the ``breaking'' point at $t=898T_L$,
for the simulation of Figs.\ref{fig:PAsim} and \ref{fig:PAsim_ErtNit}.
The peak value of $n_i$ exceeds forty times the background
value $n_0=0.005n_c$.
\label{fig:NiEr_det}}
\end{figure}

However, the ions retain
the velocity acquired during the acceleration stage. 
For $r>r_{\mbox{\tiny max}}$, 
where $r_{\mbox{\tiny max}}\simeq r_0$ is the position of the PF maximum,
the force on the ions decrease with $r$, and thus the ion final
velocity does as well. 
As a consequence the ions starting at a position $r_i(0)>r_{\mbox{\tiny max}}$
are ballistically focused towards a narrow region
at the edge of the intensity 
profile. This spatial focusing effect is actually very tight,
as in Fig.\ref{fig:PAsim} we observe a large number of ions to 
reach approximately the same position 
($r=r_b \simeq 15.5~\lambda$) at the same time 
($t=t_b \simeq 898T_L$). Here the ions pile up producing a very sharp peak 
of $n_i$. The peak value
of $n_i$ is $\simeq 46n_0=0.23n_c$ at $t=898T_L$, as shown in 
the detail of the density and field profiles of Fig.\ref{fig:NiEr_det}.
The density peak is well out of scale
of the density axis in the plots of Fig.\ref{fig:PAsim}.
Simple modeling (see \ref{app:analyticbreak}) provides an
estimate of the position $r_b$ and the instant $t_b$ at
which the density spike is formed as a function of the laser pulse
parameters:
\begin{eqnarray}
r_b \simeq \left(\frac{3}{2}\right)^{3/2}r_0
   \simeq 1.84 r_0 ,\\
t_b 
 \simeq \frac{\pi}{2}e^{3/4}\sqrt{\frac{A}{Z}\frac{m_p}{m_e}}\frac{r_0}{a_0 c}
 \simeq 10^2 T_L\sqrt{\frac{A}{Z}}\frac{r_0}{a_0\lambda}.
\end{eqnarray}
The latter expression is likely to be an underestimate for $t_b$ as the simple
model assumes the pulse intensity to be constant during the time the ions
take to reach the point $r=r_b$, while in the simulation the pulse is shorter
than $t_b$ and the value of the amplitude averaged over the pulse duration is
lower than $a_0$. For the run of Fig.\ref{fig:PAsim}
we obtain $r_b \simeq 14\lambda$ and $t_b \simeq 569~T_L$, in fair agreement
with the simulation results.

The piling up of ions at the point $r=r_b$ leads to hydrodynamical
breaking in the ion density profile, as the fastest ions 
overturn the slowest ones at $r=r_b$.
The onset of hydrodynamical 
breaking is clearly evident in the contours of $f_i$ 
at the ``breaking'' time $t_b \simeq 898T_L$ (where the typical 
``vertical'' shape of the contours of $f_i$ can be noticed).
At later times ($t=1402T_L$) the ion phase space plot is reminiscent
of the so-called ``\textsf{x}-type'' breaking that was first observed in 
the collapse of ion acoustic waves excited by Brillouin scattering
\cite{forslundPF75b}; in Ref.\cite{moriPRL88} the occurrence of 
this type of breaking at the walls of a self-focusing channel
was also mentioned, but not discussed in detail.

The ion density breaking leads to the
generation of a short ion bunch (located near $r=22\lambda$ at $t=1407T_L$),
propagating in the outward direction, 
containing nearly monoenergetic ions having $p_i \simeq 0.014 m_i c$
for the parameters of Figs.\ref{fig:PAsim}--\ref{fig:PAsim_ErtNit}. 
The motion of the bunch is almost purely ballistic, as it is evident 
in Fig.\ref{fig:PAsim_ErtNit}.

A similar feature was observed in the case of
``longitudinal'' ponderomotive acceleration described in 
Ref.\cite{macchiPRL05}, where the bunch formation is also shown
to be related to spatial focusing of ions and hydrodynamical breaking 
of the ion density. In the case investigated in Ref.\cite{macchiPRL05} 
ion acceleration is also of ponderomotive nature because the use of
circularly polarized light prevents the generation of ``fast'' electrons
and thus the onset of sheath ion acceleration.

In addition to the ion bunch, 
we observe both a small fraction of the ions which is further
accelerated at breaking (up to $\simeq 0.014 m_i c$), and another fraction 
which is accelerated inwards, having negative $p_i$ up to 
$\simeq -0.005m_i c$, and thus moves back towards the axis. 
At later times ($t\gtrsim 4000 T_L$) these ions are found to form a local
density maximum (i.e. a narrow plasma filament) around $r=0$,
as highlighted in Fig.\ref{fig:PAsim_ErtNit}. A local density maximum on
the channel axis was also found in 3D electromagnetic simulations
\cite{borghesiPRL97} but explained by magnetic pinching, which is 
absent in our model. 

The dynamics of the ions after breaking and related features, such as 
the ``fast'' ion bunch and the local density maximum on axis 
are related to the effect
of the generation of a strong ambipolar field at the breaking point, 
which we now discuss.

\subsection{Electric ``echo'' effect and electron heating}
\label{sec:echo}

In the electric field plot at the breaking time we observe 
a strong ambipolar electric field appearing around the breaking point.
The field is rather intense (its amplitude slightly exceeding that of the
positive field due to charge depletion at earlier times) and highly
transient; the complete ``movie'' of $E_r(r,t)$ in Fig.\ref{fig:PAsim_ErtNit}
shows that the field near the
breaking point rises sharply from zero to its peak value over a few laser 
cycles time, and then decreases less rapidly to lower values (see the profile 
at $t=1402T_L$). The ambipolar structure slowly moves in the outward direction 
and is observable up to very long times. The ``inversion'' of the field,
i.e. the appearance of a region where the electric fields points towards the
axis, in the direction opposite to the initial stage of electron depletion,
was evident in the proton imaging measurements reported in Ref.\cite{karNJP07}.
The electric ``echo'' is evidently correlated with the rapid and strong 
heating of electrons at breaking, which we observe in the $f_e$
plots at the breaking time. 

\begin{figure}
\begin{center}
\includegraphics[angle=0,width=0.6\textwidth]{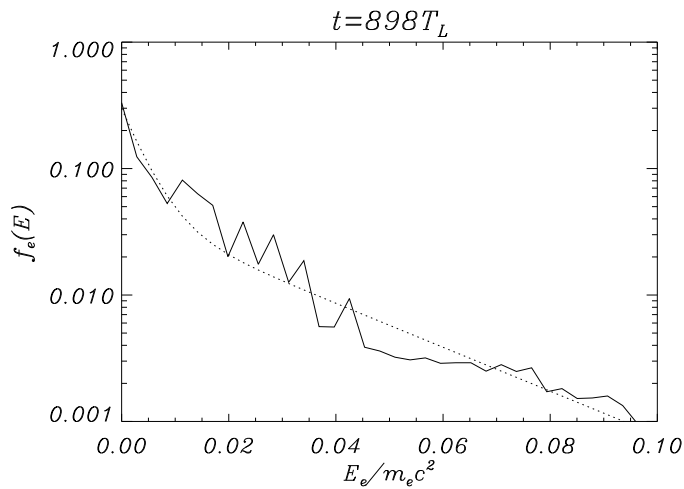}
\end{center}
\caption{Electron energy distribution at $r=15.8\lambda$ and $t=898T_L$
for the simulation of Figs.\ref{fig:PAsim}.
The dotted line is a two-temperature Maxwellian with 
$T_1=4 \times 10^{-3}m_e c^2$ and 
$T_2=2.5 \times 10^{-2}m_e c^2$.
\label{fig:feE}}
\end{figure}

The generation of an electric field around the density peak as the ions
approach the breaking point is interpreted as due to the inability of
electrons to dynamically screen the ion field when the density variation
is too fast. Let $d$ be the width of the ion peak at a certain time. 
The density changes in time due to the ballistic motion of ions
moving towards the breaking point $r=r_b$. The typical variation time 
of the ion density is $\tau_i \sim d/v_i$, where $v_i$ is the ion
velocity determined by the acceleration stage. Electrons respond on
a typical time of the order of the inverse plasma frequency, 
$t_p \sim 2\pi/\omega_p$, thus they are able to preserve plasma
neutrality if ${d}/{v_i} > t_p$. If such condition is violated,
the field around the density spike will be close to the 
unscreened field of the ions, i.e. a surface field of
amplitude $E_i \sim \pm 2\pi Zen_i d$, leading to electron heating. 
The ``neutrality condition'' can be written in terms of the 
typical units used in the paper as
\bea
\frac{(d/\lambda)}{(v_i/c)} \gtrsim \frac{2\pi}{\omega_p}
                                    ={T_L}\sqrt{\frac{n_c}{n_e}},
\label{eq:neutrality}
\eea
which is useful to check its violation in the simulation results.
However, it is noticeable that as a function of time 
$d=d(t)$ may be not independent from the density $n_e=n_e(t)$; due to 
mass conservation we expect $d \sim 1/n_i$ and thus $d \sim 1/n_e$
as far as $n_e \simeq n_i$ locally. 
Thus, as $n_e$ spikes the l.h.s. of Eq.(\ref{eq:neutrality})
decreases more rapidly than the r.h.s., boosting the violation of the
inequality. This effect might account for the ``robustness'' of the 
electric ``echo'' effect in our simulations. The correlation between 
the formation of the density spike and electron heating might also be
qualitatively understood on the basis of more general arguments, such
as the fact that small oscillations of electrons around their point
of (quasi--)equilibrium tend to become nonlinear across a sharp density
gradient, i.e. around the spike, and lead to heating.

The electron energy distribution $f_e$ near the breaking point 
at $t=898T_L$ (Fig.\ref{fig:feE}) for the simulation of of 
Figs.\ref{fig:PAsim}
may be roughly approximated by a 
two--temperatures Maxwellian 
with values $T_1 \simeq 4\times 10^{-3}m_ec^2 \simeq 2~\mbox{keV}$
and $T_2 \simeq 2.5\times 10^{-2}m_ec^2 \simeq 12.8~\mbox{keV}$.
The value of $T_1$ is 
fairly consistent with an electron acceleration in the unscreened
ion field leading to typical energies $\sim e E_i d$; in fact, 
estimating $E_i \simeq 0.05E_0$ and $d \simeq 0.1\lambda$ from 
Fig.\ref{fig:NiEr_det} we obtain a typical energy
$eE_i d \simeq 5\times 10^{-3}m_ec^2$. The presence of the ``hot'' tail 
in $f_e$ may be a signature of stochastic processes
able to accelerate a minor fraction of electrons to higher energies of the
order of $T_2$.

The electron ``temperature'' generated during the highly transient heating 
stage accounts for the persistence of an ambipolar field around the density
peak at later times. The almost Maxwellian shape of the electron energy 
distribution function allows to describe the late electric field as a 
``Debye sheath'' field generated around a thin foil of cold ions and 
thermal electrons in Boltzmann equilibrium. In such a model the foil
is ``thin'' if the spatial extension (FWHM) of the sheath field,
$\ell_s$, is larger than the extension of the density ``cusp'' $d$, which
is indeed the case in Fig.\ref{fig:NiEr_det} where
$\ell_s \simeq 0.3\lambda >d \simeq 0.1\lambda$. The model is 
analytically solvable in planar geometry (see \ref{app:thinsheath})
for a delta--like foil density, which is appropriate in
the limit $d\ll\lD$ where $\lD$ is the Debye length.
Using this analytical result for a rough estimate, 
we write the peak field $E_s$ and the sheath extension $\ell_s$ as 
\begin{equation}
E_s =2\pi e n_{i}d ,\qquad \ell_s =2\frac{4\lD^2}{d} ,
\end{equation}
where $n_{i}$ is the ion density in the foil. 
Assuming parameters values from
the 1D simulations, $n_{f}\simeq 0.23n_c$, 
$d\simeq 0.1\lambda$, we find in normalized units 
$E_s/E_0=\pi(n_f/n_c)(d/\lambda) \simeq 0.07$,
fairly consistent with the simulation results.
However, using $T_e=T_1 \simeq 4\times 10^{-3}m_ec^2$,
we obtain
$\lD^2/d=(T_e/m_ec^2)(n_c/n_f)(\lambda^2/4\pi^2d)
\simeq 4.3\times 10^{-3}\lambda$ and thus $\ell_s/\lambda \simeq 0.034$,
which is smaller than the observed value and not consistent with the 
assumption $\ell_s>d$. This suggests that the sheath is mostly formed
by the ``hot'' electrons having temperature $T_2 \simeq 6T_1$ and lower
density (roughly estimated to be $n_h \simeq 0.17 n_f$ from the 
double--Maxwellian fit of the electron distribution), so that the 
effective values of $\lambda_D^2$ and $\ell_s$ may increase by one order
of magnitude. Replacing $n_f$ by $n_h$ accounts for the screening  
by the colder electrons of the ion field acting on the ``hot'' ones. 
Due to these
effects and additional ones (e.g. the dependence of the sheath profile
on the cut-off energy for ``truncated'' Maxwellians, see 
\ref{app:thinsheath}) the simple thin sheath model cannot 
accurately describe the field around the density spike, but it is
at least useful for a qualitative description. 

For clarity it is worth to point out that
the breaking of the ion fluid and the formation of a strong ambipolar
field around the breaking point occurs also for longer
pulses, i.e. when the PF and the related electric field are not over
at the breaking time. This has been observed by varying the pulse duration in
our simulations. The electric field ``echo'' occurring when the pulse is 
already over is a remarkable signature of the strong electron heating  
that occurs following the ion fluid breaking in a regime of plasma
neutrality. On a qualitative basis, a possible explanation of the 
electron heating is that near breaking the temporal variation of the
plasma density becomes faster and a very strong density gradient is generated,
so that electron oscillations around their equilibrium positions may become
non-adiabatic, leading to heating.

The generation of the strong ambipolar field structure, creating 
a very sharp field gradient at the breaking point,
has a feedback effect on the ion distribution, acting as a sort of
``axe'' that separates slower ions from faster ones. Slower ions 
are reflected by the negative part of the field and acquire
negative radial velocity, thus driving the formation of the density 
maximum at $r=0$
at $t>1700$ in Fig.\ref{fig:PAsim_ErtNit}. Faster ions have enough energy 
to cross the electric field barrier, producing the escaping bunch, 
and may also receive additional
energy from the positive part of the field; this accounts for the 
ions which are observed to get further momentum after breaking 
in Fig.\ref{fig:PAsim} at $t=1402T_L$. These two ions components produce
the upper right and lower left ``arms'' in the \textsf{x}-structure 
of the phase space at breaking. 
As the electric field decreases very rapidly after 
breaking, the other ions form a dense front moving in radial direction 
with a velocity much smaller than the fast ions in the bunch, as shown
in Fig.\ref{fig:PAsim_ErtNit}.
Most of warm electrons remain confined around the ion front and their 
temperature is found to decrease with time.

\subsection{Transition to Coulomb explosion}
\label{sec:coulomb}

As stated above,
under the action of the PF, the electrons are pushed
away from the axis creating a back-holding electrostatic field
which balances the PF almost exactly in the ponderomotive regime.
However, the balance is possible only
if the PF does not exceed the maximum possible value of the 
electrostatic force at some radius $r$, which occurs if all electrons
have been removed up to such value of $r$, i.e. if complete electron depletion
occurs, and ions have not moved significantly yet.
Within our 1D cylindrical model, an approximate threshold condition
for complete electron depletion can be thus derived as follows.
If all electrons are removed from a central
region and the ion density is equal to its initial value, the electric
field in the depletion region is given by
\begin{equation}
E_d(r)=2\pi Ze n_{i0} r = 2\pi e n_{e0} r.
\end{equation}
If $F_r$ exceeds the force due to the ``depletion'' field $E_d(r)$, 
this will first occur near $r=0$, where $F_r$ is given approximately by
\begin{equation}
F_r \simeq \frac{m_e c^2}{r_0}\frac{r}{r_0}\frac{a_0^2}{(1+a_0^2)^{1/2}}.
\end{equation}
Hence, posing $F_r>eE_d$ for $r\rightarrow 0$ yields the condition
\begin{equation}
\frac{m_e c^2}{r_0^2}\frac{a_0^2}{(1+a_0^2)^{1/2}}>2\pi e^2 n_{e0} ,
\end{equation}
which we rewrite as
\bea
a_0>\left[{\frac{k^2}{2}+\left({\frac{k^4}{4}+k^2}\right)^{1/2}}\right]^{1/2} 
,\label{eq:threshold} \\
k=2\pi\left(\frac{e^2}{m_e c^2}\right)n_e r_0^2=2\pi r_c n_e r_0^2 ,
\eea
where 
$r_c=2.82 \times 10^{-13}~\mbox{cm}$. For $n_e=10^{19}~\mbox{cm}^{-3}$
($10^{18}~\mbox{cm}^{-3}$) and
$r_0=7.5~\mu\mbox{m}$, we find $k \simeq 10$ ($1$) and thus a complete
electron depletion is expected to occur for $a_0 \gtrsim 10$
[$\gtrsim 1.3$]. 

The onset of complete electron charge depletion near the axis may occur even 
lower intensities than predicted by Eq.(\ref{eq:threshold}) if the pulse
is not too short. In fact, even if initially $F_r =eE_r$ holds,
as the ions move under the action of the force $ZeE_r=ZF_r$ the density
near the axis must decrease, so the maximum electrostatic field that can be 
generated near the axis also decreases in time, and the condition of 
complete electron depletion may be met.

\begin{figure}
\begin{center}
\includegraphics[width=0.6\textwidth]{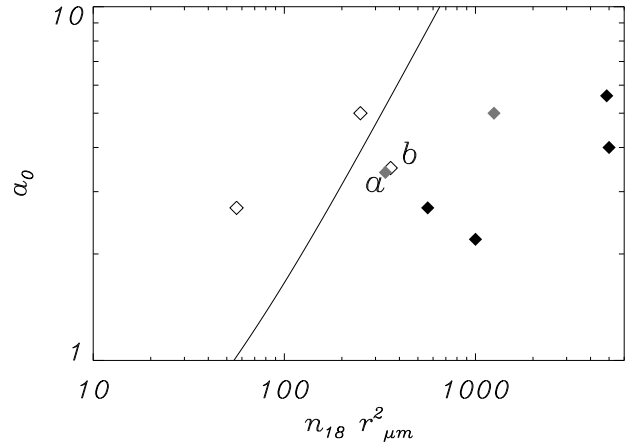}
\end{center}
\caption{Threshold for the transition from the ``ponderomotive'' to the
``Coulomb explosion'' regime.
The thick line gives amplitude threshold for complete 
electron density depletion in the central region
[Eq.(\ref{eq:threshold})] as a function
of the parameter $n_{e0}r_0^2$ with $n_{e0}$ in units of 
$10^{18}~\mbox{cm}^{-3}$ and $r_0$ in units of $1~\mu\mbox{m}$.
Filled, empty and ``gray'' dots represent simulations where 
the electron depletion is absent, strong, or ``marginal'', respectively.   
The labels $a$ and $b$ indicate two simulation for almost identical parameters
but different pulse duration, so that in the longer pulse case ($b$) electron
depletion is favored.
\label{fig:threshold}}
\end{figure}

Fig.\ref{fig:threshold} shows the threshold amplitude
as a function of $n_e r_0^2$. The data points in Fig.\ref{fig:threshold}
represent the values of $n_e r_0^2$ for numerical simulations where 
electron depletion in the central region is either absent (empty, ``white'' 
diamonds) or evident (filled, ``black'' diamonds). With ``gray'' diamonds
we indicate ``near--threshold'' cases where depletion occurs just very near
to the $r=0$ axis and the maximum force on ions is still given by $ZF_r$,
i.e. the PF is larger than $E_r$ over most of the spatial range.
The distribution of data points confirm that Eq.(\ref{eq:threshold}) gives
just a rough condition for the transition from the ponderomotive to 
the Coulomb explosion regime and that this transition also depends on 
the pulse duration. For example, the labels \emph{a} and \emph{b} in 
Fig.\ref{fig:threshold} indicate two simulations which have very similar
values of $a_0$ and $n_e r_0^2$, but for \emph{b} the electron depletion
is much stronger. This can be explained by the pulse duration for case 
\emph{b} that is roughly two times the value for case \emph{a}.

\begin{figure}
\begin{center}
\includegraphics[angle=0,width=0.8\textwidth]{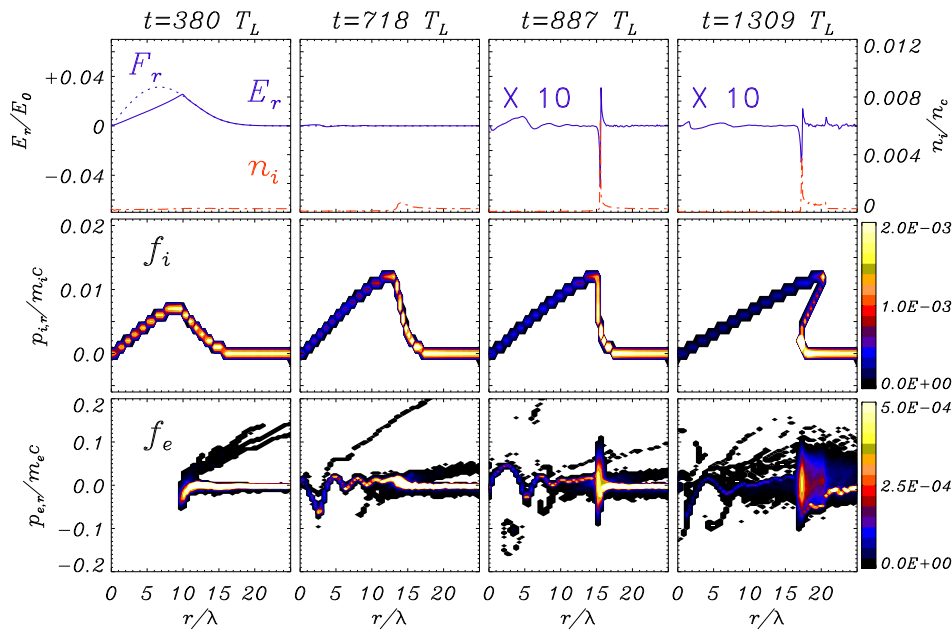}
\end{center}
\caption{(color online) Simulation results in the regime of strong electron
depletion and Coulomb explosion. 
All the fields and distributions are the same of Fig.\ref{fig:PAsim},
except that the normalized 
ponderomotive force $F_r/e$ is shown (blue, dotted line) 
in the leftmost upper plot.
The vertical scales for the $E_r$ and $p_{i,r}$ (ion momentum) are the same 
of Fig.\ref{fig:PAsim}, while to
improve readability in the upper two plots for $t=902T_L$ and $t=1402T_L$, 
$E_r$ has been multiplied by $10$, and the scale for $p_{e,r}$ has been 
shrunken with respect to Fig.\ref{fig:PAsim}.
Parameters are 
$a_0=2.7$, $n_e/n_c=10^{-3}$, $r_0=7.5\lambda$, $\tau_{1/2}=330T$.
\label{fig:TRsim}}
\end{figure}

In the Coulomb explosion regime, the ions in the 
region of electron depletion will be accelerated by their own
space charge field. 
Fig.\ref{fig:TRsim} shows results of a 1D simulation where the laser pulse
parameters are the same of Fig.\ref{fig:PAsim}, but the density has been
lowered by a factor of 10 (thus, $n_e/n_c=10^{-3}$). For such parameters, from 
Eq.(\ref{eq:threshold}) we expect to enter the Coulomb explosion regime.
The plot of $E_r$ and of the scaled PF $F_r/e$ in Fig.\ref{fig:TRsim}
at $t=380T_L$ and the corresponding contour plot of $f_e$
show that complete charge depletion occurs in the 
region $r<r_d \simeq 8~\lambda$. For $r>r_d$, $eE_r \simeq F_r$ still 
holds. We notice that at the boundary of the depletion layer 
($r \simeq r_d$), where
the force balance breaks down, a few electrons are accelerated to relatively 
high velocity and escape towards the outer region.

Due to electron depletion the resulting maximum force on the ions 
$F_{\mbox{\tiny max}} \simeq Ze E_r(r_d)$
is less than the maximum of $ZF_r$; thus, the maximum momentum of ions 
at the end of the laser pulse ($t \simeq 718T_L$) is lower than in the
case of Fig.\ref{fig:PAsim} for the same value of the laser intensity.
Hence, for a given laser pulse, ion acceleration saturates for decreasing
plasma densities as the condition (\ref{eq:threshold}) is crossed. 
The plots corresponding to the later times show that the spiking 
and breaking of the ion density, followed by ion bunch formation and 
ambipolar field generation, still occurs for these parameters; however,
both the field amplitude and the electron energy are lower now. 
In simulations where the electron density is further lowered, 
electron depletion occurs over almost all the pulse profile,
almost all ions are accelerated by the Coulomb explosion, while
bunch formation and ambipolar field generation tend to disappear.

\subsection{Saturation mechanisms for ion acceleration}
\label{sec:saturation}

In the ponderomotive regime as defined above, the force on the 
ions is given by $ZeE_r=-Zm_e c^2\partial_r [1+a^2(r,t)]^{1/2}$.
If the temporal dependence of the pulse intensity is neglected,
i.e. $a^2(r,t) \simeq a^2(r)$, one may use the 
``ponderomotive potential'', $\Phi_p(r)=Zm_e c^2[(1+a^2(r))^{1/2}-1]$
to estimate the final energy of ions as a function of their starting
position. This assumption leads to a maximum ion energy equal to the
peak value of $\Phi_p$,
\begin{equation}
U_i^{\mbox{\tiny max}} \simeq \Phi_p(r=0) 
\simeq Zm_e c^2 \left[(1+a_0^2)^{1/2}-1\right],
\label{eq:Umax_K}
\end{equation}
that is the energy acquired by ions initially located on the 
channel axis, i.e. at $r=0$. In Ref.\cite{krushelnickPRL99} this
relation has been actually used to estimate the peak intensity on axis from
the analysis of the ion spectrum.

A clear limitation of this simple picture is the quite long time it would
take for an ion starting at $r=0$ to go downhill the ponderomotive 
potential and to gain the maximum energy, as the PF is very small near
$r=0$. Thus, the laser pulse duration may be shorter than the acceleration
time, limiting the ion energy. This is indeed the case for the simulation 
of Fig.\ref{fig:PAsim} where the laser pulse is over before the fastest ions
have moved out of the radial extension of the pulse. The maximum ion energy
observed in Fig.\ref{fig:PAsim} is about four times lower than the 
estimate based on Eq.(\ref{eq:Umax_K}). In this case using the ion energy
cut-off to evaluate $a_0$ in Eq.(\ref{eq:Umax_K}) would lead to an
underestimate of the peak intensity.

As discussed in section \ref{sec:coulomb}, the possible onset of complete 
electron depletion is another limiting factor for the ion energy as 
it gives an an upper limit to the accelerating force. The overall energy
spectrum may also be modified by the onset of hydrodynamical breaking
at the edge of the beam radius as discussed in sections 
\ref{sec:breaking} and \ref{sec:echo}.

\begin{figure}
\begin{center}
\includegraphics[angle=0,width=0.8\textwidth]{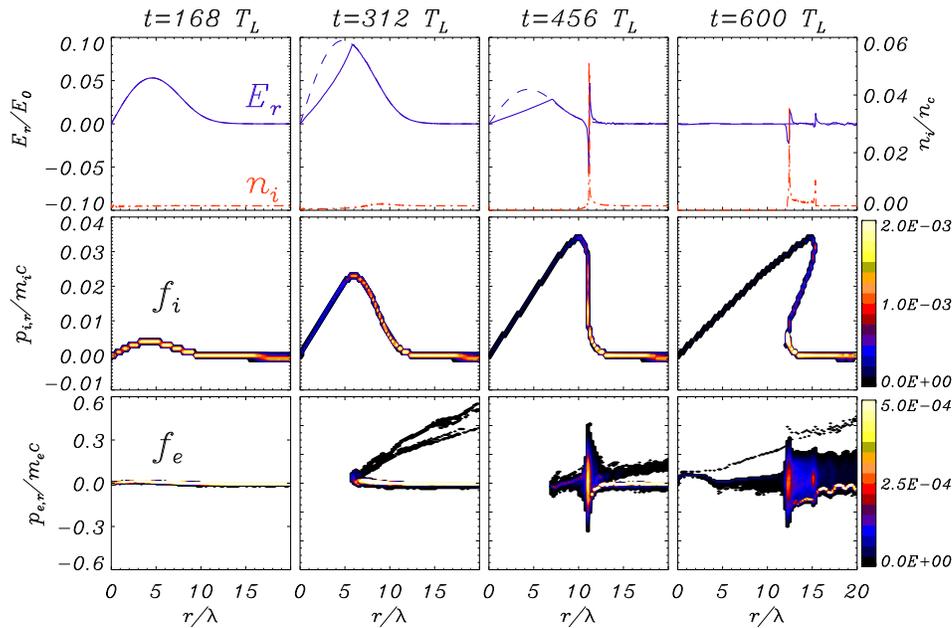}
\end{center}
\caption{(color online) Simulation results for parameters similar to those
of Ref.\cite{krushelnickPRL99} 
($a_0=3.5$, $n_e/n_c=0.01$, $r_0=5\lambda$, $\tau_{1/2}=300T$).
All the fields and distributions are the same of Figs.\ref{fig:PAsim}
and \ref{fig:TRsim}.
\label{K99}}
\end{figure}

To compare with preceding work we have simulated a case with almost the
same parameters as those used in the calculations performed in 
Ref.\cite{krushelnickPRL99} to interpret
the experimental data. A He plasma is considered and the parameters are 
$n_e/n_c=0.01$, $r_{1/2} \simeq 1.2r_0=6\lambda$, $\tau_{1/2}=300T$, 
and $a_0=3.5$ [this is lower than the value $a=5$ quoted in 
Ref.\cite{krushelnickPRL99} due to a different convention in the expression
of the PF, i.e. $F_r=(1+a^2/2)^{1/2}$].
Simulation results are shown in Fig.\ref{K99}. The maximum ion energy,
corresponding to the ions in the ``fast'' bunch, is $\simeq 2.2~\mbox{MeV}$,
very close to the energy spectrum cut-off in Fig.2 of 
Ref.\cite{krushelnickPRL99}, and significantly lower than the 
$2.7~\mbox{MeV}$ value corresponding to Eq.(\ref{eq:Umax_K}).
This is due to the complete electron depletion in the central region,
occurring about 60 cycles before the pulse peak and keeping
the electrical force $eE_r$ below the maximum of the PF $F_r$, 
as shown in Fig.\ref{K99}. The breaking of the ion density profile occurs 
at $r \simeq 11~\lambda$, very close to the analytical estimate given by 
Eq.(\ref{eq:rb}), and at $t \simeq 456T_L$, roughly two times the 
value obtained from Eq.(\ref{eq:tb}).

\section{Discussion and conclusions}

The one-dimensional electrostatic, ponderomotive model used in this paper
to investigate the dynamics of self-channeling
yields results whose agreement with 
experiments is remarkable, taking the simplicity of the model into
account. A prominent example has been provided by proton imaging data 
which have been reproduced by simulating the particle deflection in the 
electric field computed by the present model \cite{karNJP07}.
Ion spectra have been also calculated for different laser and plasma 
parameters and agree with measurements within experimental uncertainties.

Of course one should not forget that the model gives a very simplified
description of the laser-plasma interaction, neglecting effects such as
pulse diffraction and nonlinear evolution, and so on. The cases in which
the simple model is successful in reproducing quantitatively or, at least,
qualitatively some features observed in experiments or in more 
complex and self-consistent simulations, correspond to regimes in which
the plasma dynamics during and after the self-channeling of the laser
pulse is dominated by ponderomotive and electrostatic forces, other 
effects playing a secondary role. An example has been provided by the 
2D electromagnetic simulations reported in Ref.\cite{macchiPPCF07}
where the breaking of the channel walls, which has been characterized in
detail with the 1D model, causes the formation of two secondary laser
beamlets propagating at an angle $\sim\lambda/r_0$, consistently with
a ``leaking waveguide'' picture. 

The analysis of the simulation results has evidenced details of the 
dynamics of ion acceleration and electric field generation. In particular,
hydrodynamical breaking has been shown to play an important role,
causing electron heating, formation of an ambipolar field around the density
cusp and, finally, affecting the ion spectrum. Unfolding this dynamics
provided an insight on the formation of ``{\sf x}-type'' structures
in the ion phase space which had been previously observed in different
contexts \cite{forslundPF75b}. The related production of a dense, 
quasi-monoenergetic ``bunch'' of ions revealed similarities with the 
process of ponderomotive acceleration of ions in overdense plasmas 
\cite{macchiPRL05}. A prominent consequence of hydrodynamical breaking,
occurring for pulse durations shorter than the breaking time, is the 
``echo'' observed in the electric field, i.e. its sudden rebirth after 
having disappeared at the end of the laser pulse.

Throughout our work we found useful to distinguish the regime of
``ponderomotive acceleration'' from that of ``Coulomb explosion''.
In the first case, complete depletion of the electron density  
does not occur, the ponderomotive and the electric forces balance 
almost exactly and electrons are in a state close to mechanical equilibrium 
at all times before breaking; in such a regime the ponderomotive potential
can be used to estimate the ion energy given the laser intensity 
(or vice versa), although the effects of finite pulse duration must be
considered. In the second case, electron density depletion occur near the
axis and the ions in this region are accelerated by their own space-charge
field, leading to a saturation of the peak 
ion energy versus the laser intensity. An approximate analytical criterion 
for the transition between the two regimes have been given and tested by
simulations, which also shows that such transition occurs smoothly.

\ack
This work has been supported by the Royal Society (UK) via a Joint
Project, by the Ministry of University and Research (Italy) via a PRIN project
and by CNR (Italy) via a RSTL project. A.~M. acknowledges 
Queen's University, Belfast, UK for an International Fellowship.
We are grateful to Alessandra Bigongiari, Peter Mulser, Francesco Pegoraro
and Vladimir T. Tikhonchuk for useful comments and discussions.

\appendix

\section{Analytical estimate of breaking point location}
\label{app:analyticbreak}

A simple model can be used
to account for the spatial focusing effect and estimate its
characteristic parameters (i.e. the position and the instant at
which the density spike is formed).
For the sake of simplicity, let us neglect 
the temporal variation of $F_r$.
If $F_r$ was a linear function of $r$,
the ion equation of motion would be of the harmonic oscillator type,
\begin{equation}
m_i\frac{d^2 r_i}{dt^2} = ZF_r \simeq -k(r_i-r_b) ,
\label{eq:harmonic}
\end{equation}
implying that \emph{all} the ions starting from an arbitrary radius
$r_i(0)<r_b$ would get to $r=r_b$ at the time 
$t_b=(2\pi/\Omega)/4$, where $\Omega\equiv \sqrt{k/m_i}$. 
In our case $F_r$ is not a linear function of $r$, 
but a linear approximation
of $F_r$ is quite accurate
around the point $r=r_f$ such that
$\partial_r^2 F(r)|_{r=r_f}=0$. 
We thus estimate the parameters 
$r_b$ and $k$ in Eq.(\ref{eq:harmonic}) from such a linear approximation.
To further simplify the derivation, 
we take the non--relativistic approximation
and write
\begin{equation}
F_r \simeq Zm_e c^2 \partial_r ({a^2(r)/2}) ,\qquad
a(r)=a_0 e^{-r^2/2r_0^2} .
\end{equation}
The non-relativistic approximation turns out not to be very bad 
even if $a_0 \simeq 1$ because at $r=r_f>r_0$ 
the exponential factor is already small.
By differentiating $F_r$ for two times, we easily obtain 
$r_f=\sqrt{3/2}r_0$. Expanding in Taylor's series around $r=r_f$ 
we obtain
\begin{eqnarray}
k&=&F'(r_f)=\frac{Z m_e c^2 a_0^2}{m_i r_0^2}e^{-3/2} , \\
r_b&=&r_f+\frac{F(r_f)}{k}=\frac{3}{2}r_f =\left(\frac{3}{2}\right)^{3/2}r_0
   \simeq 1.84 r_0 .
\label{eq:rb}
\end{eqnarray}
Thus, $r_b$ depends on $r_0$ only, and for the run of Fig.\ref{fig:PAsim}
we obtain $r_b \simeq 14\lambda$.
From the breaking time we obtain
\begin{equation}
t_b =\frac{\pi}{2}\sqrt{\frac{k}{m_i}}
       =\frac{\pi}{2}e^{3/4}\sqrt{\frac{A}{Z}\frac{m_p}{m_e}}\frac{r_0}{a_0 c}.
\label{eq:tb}
\end{equation}
This is likely to be an underestimate for the breaking time, 
since in our case the pulse duration is shorter than the 
ion acceleration time and it may be proper to replace the peak value of 
the amplitude $a_0$ with some time--averaged value
$\bar{a}<a_0$. 

The predictions of the rough linear approximation of $F_r$ are thus in
fair agreement with the simulation results. The important point to stress
is that in the ponderomotive regime $r_b$ depends on $r_0$ only, 
while $t_b$ depend only on the laser pulse parameters and on the
$Z/A$ ratio, but not on the plasma density.
Our numerical simulations performed for different parameters 
in this regime 
show that the spatial focusing and piling-up of ions is a robust 
phenomenon which, once the spatial form
of $F_r$ is fixed, tends to occur always at the same point.

It is worth to notice that these estimates have been obtained for a Gaussian
intensity profile. A different functional form would produce different
results for $r_b$ and $t_b$, but their scaling with pulse width and amplitude
should be the same. In general we expect any reasonable ``bell--shaped''
profile of the laser pulse to produce a sharp density increase near the 
edge of the beam, as this is the result of the decreasing with radius 
of the ponderomotive force in such region. The spiking of the density is sharp
for ponderomotive force profiles such that $\partial_r^2 F(r)=0$ at some
point.

\section{The sheath field around a thin foil}
\label{app:thinsheath}

In this section we compute 
analytically the sheath field around a plasma foil 
having a thickness much less than the Debye length,
using a one--dimensional, cartesian geometry. 
The plasma foil is modeled as a delta--like distribution with the ion
density given by
\beq
n_i(x)=n_0 d\delta(x),
\eeq
where $n_0 d$ is the surface density of the foil. Ions are supposed to 
be fixed.
The electrons are assumed to be in Boltzmann equilibrium
\beq
n_e={n_0}\exp\left(-\frac{U}{T_e}\right)
   ={n_0}\exp\left(\frac{e\Phi}{T_e}\right),
\eeq
where $\Phi$ is the electrostatic potential. 
It is convenient to express the coordinate, the potential
and the electric field in dimensionless form,
\beq
z=\frac{x}{\lD},\qquad \phi=\frac{e\Phi}{T_e},
\qquad \varepsilon = \frac{e E}{T_e\lD},
\eeq
where ${\lD^2}={T_e}/{(4\pi e^2 n_0)}$ is the Debye length.
From Poisson's equation we thus obtain
\beq
\phi''=e^{\phi} .
\eeq
We expect a potential that will be a even function of $z$ (so that 
$n_e$ is even and $E$ is odd), thus we can restrict the analysis 
to the $z>0$ region.

\begin{figure}
\begin{center}
\includegraphics[width=0.6\textwidth]{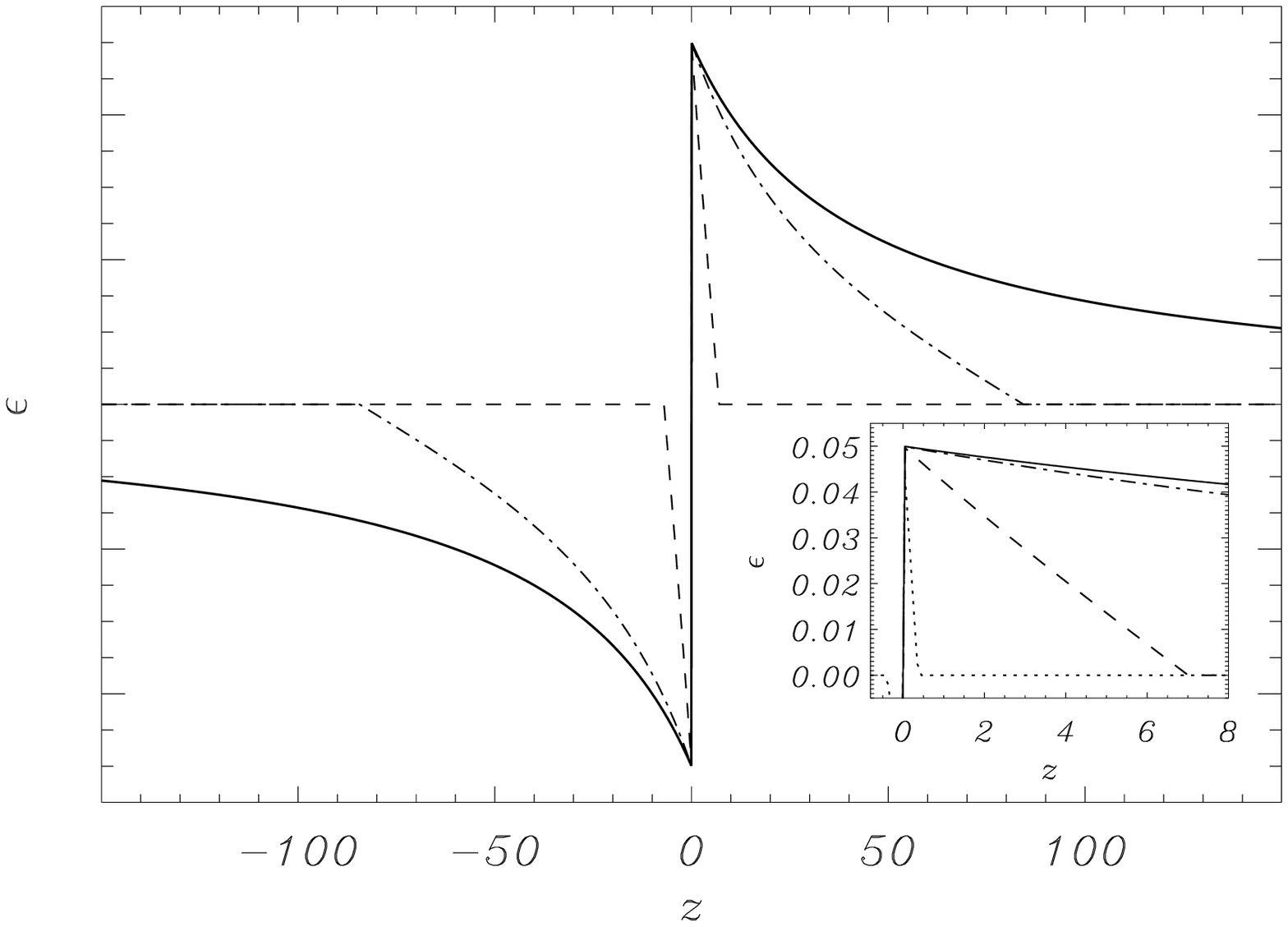}
\end{center}
\caption{The field profile in the case $d/\lD=0.1$
for a cut--off energy $u=\infty$ (thick line), 
$u=5$ (dashed), and $u=8$ (dashed-dotted).
The inner plot shows the profiles for small positve values of $z$
including the $u=2$ case (dotted).
\label{fig:thinfield}}
\end{figure}

Multiplying by $\phi'$ and integrating once we obtain
\beq
\frac{1}{2}[\phi'(z)]^2-\frac{1}{2}[\phi'(0^+)]^2
=e^{\phi(z)}-e^{\phi(0)}.
\label{eq:BC}
\eeq
Here $\phi'(0^+)\equiv -\varepsilon_0$ is the electric field at the surface.
If the system is neutral, the field at $z=+\infty$ is zero, thus
\beq
-\frac{1}{2}\varepsilon_0^2=-\frac{1}{2}[\phi'(0^+)]^2
=e^{\phi(+\infty)}-e^{\phi(0)}.
\eeq
Our first ansatz is to take 
$e^{\phi(+\infty)}=0$,
from which we obtain 
$[\phi'(0^+)]^2/2-e^{\phi(0)}=0,$
and Eq.(\ref{eq:BC}) becomes
\beq
\frac{1}{2}[\phi'(z)]^2=e^{\phi(z)}.
\eeq
This latter equation can be integrated by the substitution
$\phi=\ln(f)$, i.e. $f(z)=e^{\phi(z)}$. 
We obtain the potential
\beq
\phi(z)=\ln\left[\frac{2}{(z+2/\varepsilon_0)^2}\right]
       =-2\ln\left(z/2+1/\varepsilon_0)\right),
\eeq
and the electric field
\beq
\varepsilon(z)=-\phi'=\frac{1}{z/2+1/\varepsilon_0} .
\eeq
From Gauss's theorem we have $E_0=2\pi e n_0 d$. 
Noting that 
$\varepsilon_0=\frac{1}{2}\frac{d}{\lD}$,
we finally find
\beq
\varepsilon(z)=-\phi'=\frac{2}{|z+4/\hat{d}|} ,
\eeq
where $\hat{d}={d}/{\lD}$ is the foil thickness 
in units of the Debye length.
This solution is shown in Fig.\ref{fig:thinfield}.

In dimensional units
and for the whole sheath
\beq
E(x)=\frac{2T_e/e}{|x|+4\lD^2/d}\mbox{sign}(x)
    =\frac{2\pi e n_0}{1+|x|d/4\lD^2}\mbox{sign}(x).
\eeq
It is worth to notice that in this solution 
the peak field, $E_0=E(x=0^{\pm})$, does
\emph{not} depend on $T$, and that the spatial extension of the
sheath (i.e. the distance at which the electric field falls by a factor
$1/2$), given by
\beq
L=\frac{4\lD^2}{d},
\eeq
is inversely proportional to $d$: the thinner the foil, the larger the sheath.
Our approach is valid if $L\gg d$, i.e. $\lD\gg d$.

In the above solution, the fact that the field extends up to infinity is
related to the energy spectrum of electrons extending up to infinite
energies. In some situation it may be more physically meaningful to
assume that the electrons have a maximum energy $U_{\mbox{\tiny max}}=uT_e$
and ``truncate'' the maxwellian up to such cut--off energy.
Pursuing this latter approach, we restart from Eq.(\ref{eq:BC}). The maximum 
of the potential energy will be equal to $U_{\mbox{\tiny max}}=uT_e$, i.e. 
the maximum of $\phi$ is $-u$, and $\phi$ will be constant beyond that
point. Thus, Eq.(\ref{eq:BC}) now reads
\beq
-\frac{1}{2}{\varepsilon^2_0}=-\frac{1}{2}[\phi'(0^+)]^2
=e^{-u}-e^{\phi(0)},
\label{eq:BC2}
\eeq
and thus the equation for $\phi$ becomes
\beq
\phi'=-\sqrt{2(e^{\phi}-e^{-u})}.
\eeq
Using the usual substitution $f=e^{\phi}$ 
we obtain after some algebra
\beq
\phi=-u+\ln\left\{
     1+\tan^2\left[{\frac{e^{-u/2}}{\sqrt{2}}}z
                   -\arctan(\frac{\varepsilon_0}{\sqrt{2}}e^{u/2})
      \right]\right\}.
\eeq
The cut--off occurs at the point $z_r$ where $\phi=-u$, i.e. 
when the argument of the $\tan^2$ function equals zero; we thus find 
\beq
z_r=\sqrt{2}e^{u/2}\arctan(\frac{\varepsilon_0}{\sqrt{2}}e^{u/2}) .
\eeq
The electric field is given by
\beq
\varepsilon(z)=\sqrt{2}{e^{-u/2}}\tan\left[-{\frac{e^{-u/2}}{\sqrt{2}}}z
                   +\arctan(\frac{\varepsilon_0}{\sqrt{2}}e^{u/2})
      \right],
\eeq
and $\varepsilon(z_r)=0$. 
The boundary conditions at $z=0$ remain the same, thus 
$\varepsilon_0={d}/{2\lD}$.

Fig.\ref{fig:thinfield} shows the field profile for different values of
the cut--off energy $u$. The sheath extension $z_r$ is strongly dependent 
on $u$.

\section*{References}

\bibliographystyle{unsrt}

\bibliography{paper}

\end{document}